\def\draftversion{false}
\RequirePackage{ifthen}
\ifthenelse{\equal{\draftversion}{true}}{
  \documentclass[aps,prb,10pt,galley,amsmath,amssymb,showpacs,
                superscriptaddress]{revtex4}
}{
  \documentclass[aps,prb,10pt,twocolumn,amsmath,amssymb,showpacs,
                 superscriptaddress]{revtex4-1}
}

\usepackage{times}
\usepackage{amsmath}
\usepackage{amssymb}
\usepackage{graphicx}
\usepackage[usenames]{color} 
\usepackage{bm} 
\usepackage{subfigure}
\newcommand{\angstrom}{\text{\normalfont\AA}}

\def\Z2{$\mathbb{Z}_2$}

\def\bise{Bi$_2$Se$_3$}
\def\bite{Bi$_2$Te$_3$}
\def\inse{In$_2$Se$_3$}
\def\sbse{Sb$_2$Se$_3$}
\def\I{\uppercase\expandafter{\romannumeral 1}}
\def\II{\uppercase\expandafter{\romannumeral 2}}
\def\III{{\uppercase\expandafter{\romannumeral 3}}}
\def\IV{{\uppercase\expandafter{\romannumeral 4}}}
\def\V{{\uppercase\expandafter{\romannumeral 5}}}

\def\Pt{\widetilde{P}}
\def\Qt{\widetilde{Q}}

\def\ket#1{\vert#1\rangle}
\def\bra#1{\langle#1\vert}
\def\ip#1#2{\langle#1\vert#2\rangle}

\def\Tra#1{\textrm{Tr}[#1]}

\def\kk{\mathbf{k}}
\def\rr{\mathbf{r}}
\def\GG{\mathbf{G}}
\def\RR{\mathbf{R}}

\def\bnk{_{n\kk}}
\def\bmk{_{m\kk}}

\def\bocc{_\mathrm{occ}}
\def\bso{_\mathrm{so}}
\def\bR{_\mathrm{R}}

\def\ut{\tilde{u}}

\def\pt{\tilde{\psi}}
\def\Ct{\widetilde{C}}

\begin{document}

\title{Spin-orbit spillage as a measure of band inversion in insulators}
\date{today}

\author{Jianpeng Liu}
\affiliation{ Department of Physics and Astronomy, Rutgers University,
 Piscataway, NJ 08854-8019, USA }

\author{David Vanderbilt}
\affiliation{ Department of Physics and Astronomy, Rutgers University,
 Piscataway, NJ 08854-8019, USA }

\date{\today}

\begin{abstract}
We propose a straightforward and effective approach for quantifying
the band inversion induced by spin-orbit coupling in band
insulators.  In this approach we define a quantity as
a function of wavevector in the Brillouin zone measuring
the mismatch, or ``spillage'',
between the occupied states of a system with and
without SOC.  Plots of the spillage throughout the BZ provide
a ready indication of the number and location of band inversions
driven by SOC.  To illustrate the method, we apply this approach to
the two-band Dirac model, the 2D Kane-Mele model, 
a 2D Bi bilayer with applied Zeeman field,
and to first-principles calculations of some
3D materials including both trivial and \Z2\ topological
insulators. We argue that the distribution of spillage in the BZ is
closely related to the topological indices in these systems.
Our approach provides a fresh perspective for understanding
topological character in band theory, and should be helpful in
searching for new materials with non-trivial band topology.
\end{abstract}


\maketitle


\def\scr{\scriptsize}
\ifthenelse{\equal{\draftversion}{true}}{
  \marginparwidth 2.7in
  \marginparsep 0.5in
  \newcounter{comm} 
  \def\commnext{\stepcounter{comm}}
  \def\commtext{{\bf\color{blue}[\arabic{comm}]}}
  \def\commmar{{\bf\color{blue}[\arabic{comm}]}}
  \def\dvm#1{\commnext\marginpar{\small DV\commmar: #1}\commtext}
  \def\jlm#1{\commnext\marginpar{\small JPL\commmar: #1}\commtext}
  \def\mlab#1{\marginpar{\small\bf #1}}
  \def\tnewpage{\marginpar{\small Temporary newpage}\newpage}
}{
  \def\dvm#1{}
  \def\jlm#1{}
  \def\mlab#1{}
  \def\tnewpage{}
}

\section{Introduction}
\label{sec:intro}

Spin-orbit coupling (SOC) is a relativistic effect originating
from the interaction
between the spin and orbital motions of electrons.
It has played a key
role in various aspects of condensed-matter physics, including the
electronic structure of solids and the
transport properties in mesoscopic systems.
\cite{winkler2003,nikolic2010oxford} It has been known since the
1950s that SOC can induce anisotropic spin splitting
in some \III-\V\
semiconductors with the zinc-blende structure,
known as the Dresselhaus splitting.\cite{winkler2003}
In 2D and quasi-2D
systems, the SOC resulting from the electric
field perpendicular to the 2D plane
gives rise to a Rashba splitting linear in $k$ with
interesting ``helical'' spin textures.\cite{winkler2003,nikolic2010oxford}
The SOC is also crucial in determining the transport behavior of
low-dimensional electronic systems. One famous example
is the weak antilocalization
in spin-orbit-coupled 2D electronic systems,
where the backscattering
amplitudes interfere destructively due to a geometric Berry
phase\cite{berry84} associated with the intrinsic SOC,
leading to a suppressed resistivity when an
external magnetic field is absent.\cite{hln}
SOC is also responsible for
spin precession in 1D and quasi-1D
systems,\cite{nikolic2010oxford} the spin Hall effect in paramagnetic
metals,\cite{Hirsch99} and numerous other effects.

The SOC has received renewed attention recently because of its
central role in the physics of topological insulators (TIs) and
related topological states.  Typically, the transition from a
topologically trivial to a non-trivial phase is accomplished by
a SOC-driven inversion of states of different symmetry at the
conduction-band minimum (CBM) and valence-band maximum (VBM).
For example, such a SOC-driven topological band inversion
between $\Gamma_6$-derived ($s$-like) and $\Gamma_8$-derived
($p$-like) states at the zone center is responsible for the
quantum spin Hall (QSH) state observed in HgTe/CdTe quantum
wells.\cite{bhz06,konig06science}
Similarly, the Kane-Mele model of 2D graphene-like systems
\cite{kane-prl05-a,kane-prl05-b} enters the QSH state
when two band inversions occur at the K and K$'$ points as the
SOC strength is increased at a constant staggered potential.
In 3D band insulators with time-reversal (TR) symmetry,
a SOC-induced band inversion can transform the system from a
trivial insulator into a strong TI displaying an odd number of
gapless Dirac cones in the surface states, as occurs for
\bise\ and \bite.\cite{kane-rmp10,zhang-rmp11,yao-review12,zhang-np09}

In the case of a 3D strong TI with inversion symmetry such as
\bise, the strong \Z2\ index can be
uniquely determined by the parities of the occupied bands
at the TR-invariant momenta (TRIM) in the Brillouin zone
(BZ).\cite{fu-prb07} If the highest occupied states and lowest
unoccupied states at one of the TRIM possess opposite parities
without SOC, and they are inverted by turning on SOC, then the
system transforms from a normal to a topological insulator.
For example, in \bise, two pairs of Kramers-degenerate occupied states
at the BZ center ($\Gamma$) are inverted by SOC, resulting in
the nontrivial \Z2\ index.  For TIs without inversion symmetry,
the band inversion may happen at arbitrary points in the BZ,
instead of at the TRIM.  We can identify such band inversion
points as the points where a band touching occurs between valence
and conduction bands as the SOC is adiabatically turned on; TR
symmetry implies that an inversion at $\mathbf k_0$ will always
be accompanied by one at $-\mathbf k_0$.
Even in the absence of inversion
symmetry, therefore, a band inversion driven by SOC is typically
a hallmark of the non-trivial topology in TIs with TR symmetry.

The SOC also plays a crucial role in giving rise to the Chern insulator
(CI) state, also known as the quantum anomalous Hall state,
which can occur in 2D insulators lacking time-reversal symmetry.
The possibility of a CI state was first introduced by
Haldane,\cite{Haldane-model} who constructed an explicit model
that demonstrates the effect.  Although the Haldane model is a model
of spinless Fermions on a honeycomb lattice, its key feature is the
presence of complex second-neighbor hoppings, which can be regarded
as arising from intrinsic atomic SOC through a second-order
perturbation process in a more realistic system of spinor
electrons.\cite{hongbin-prb13} An example is a Bi bilayer with an applied 
Zeeman field, as will be discussed below.

The concept of topological band inversion
has been much discussed in the topological-insulator literature,
but in the absence of symmetry it may be difficult to recognize
when a band inversion has actually occurred.  The usual approach is to
look at the symmetry or orbital character at a high-symmetry
point in the BZ where a band inversion is suspected, but this
only works if sufficient symmetry is present.  Some authors
have tried to deduce the presence of band-inversion behavior by
studying other properties of the system, such as by looking at
the qualitative shape of the bands near the symmetry point,
\cite{klintenberg-archive10} or even more indirectly, by studying
the variation of the band-energy differences with
strain in the absence of SOC.\cite{yang-nm12} However, the
reliability of such methods is questionable, as they do not
give a direct and quantitative evaluation of the SOC-induced
band inversion.

In this paper, we propose that the calculation of spin-orbit
spillage, which measures the degree of mismatch between the
occupied band projection operators with and without SOC, provides
a simple and effective measure of SOC-driven band inversion
in insulators.  We demonstrate that the mapping of this spin-orbit spillage
in $k$-space easily allows a direct identification of any region
in the BZ where band inversion has occurred, and that the maximum
spillage is a useful indicator of topological character.
We illustrate the method in the context of both tight-binding
models and realistic first-principles calculations.

The paper is organized as follows. In Sec.~\II\ the
formal definition of SOC-induced spillage is introduced, and
the correspondence between topological indices
and spillage is also discussed.
In Sec.~\III\ the formalism is applied to various
systems, including the two-band Dirac model, 
2D Kane-Mele model, a Bi bilayer
with tunable SOC and exchange field, and realistic materials
including \bise, \inse, and \sbse.
In Sec.~\IV\ we make a brief summary.

\section{Formalism}
\label{sec:formal}

\subsection{Definitions}
\label{sec:def}

Mathematically, the mismatch between two projection operators $P$
and $\Pt$, both of rank $N$, can be represented by a quantity
\begin{equation}
\gamma=N-\Tra{P\Pt}=\Tra{P\Qt}=\Tra{Q\Pt}
\label{equa:spillage}
\end{equation}
where $Q=1-P$ and $\Qt=1-\Pt$ denote the complementary projections.
This measure of mismatch is often referred to as ``spillage''
since it measures the weight of states that spill from $P$ into $\Qt$,
or equivalently, from $\Pt$ into $Q$.
Clearly the spillage vanishes if $P=\Pt$ at one extreme, and rises
to $N$ at the other extreme if there is no overlap at all
between the subspaces associated
with $P$ and $\Pt$.  Thus, the spillage provides a measure of
the degree of mismatch between the two subspaces.

Here we apply this concept to the band projection operators
\begin{equation}
P(\kk)=\sum_{n=1}^{n\bocc} \ket{\psi\bnk}\bra{\psi\bnk}
\label{equa:bandproj}
\end{equation}
associated with a given wavevector $\kk$ in the BZ
of a crystalline insulator with $N\!=\!n\bocc$ occupied bands.
We assume
an effective single-particle Hamiltonian such as that appearing
in density-functional theory (DFT).\cite{dft1,dft2}
Then the SOC-induced spillage $\gamma(\kk)$ is defined as
\begin{equation}
\gamma(\kk)=\Tra{P(\kk) \Qt(\kk)}
\label{equa:def1}
\end{equation}
where $P$ and $\Pt$ (and their complements) refer to the
system with and without SOC respectively.
More explicitly,
\begin{align}
\gamma(\mathbf{k})& =n\bocc-\Tra{P(\mathbf{k})
\Pt(\mathbf{k})}\notag\\
& =n\bocc-\sum_{m,n=1}^{n\bocc}\vert M_{mn}(\kk)\vert^2
\label{equa:def2}
\end{align}
where
\begin{equation}
M_{mn}(\kk)= \ip{\psi\bmk}{\pt\bnk}
\label{equa:Mmn}
\end{equation}
is the overlap between occupied Bloch functions with and without SOC at the
same wavevector $\kk$.
Equivalently, this can be written as $M_{mn}(\kk)= \ip{u\bmk}{\ut\bnk}$
if one prefers to work in terms of the cell-periodic $\ket{u\bnk}$
defined as $u\bnk(\rr)=e^{-i\kk\cdot\rr}\psi\bnk(\rr)$.

In the case of realistic DFT calculations in a plane-wave basis,
the overlap matrix elements
are easily evaluated as
\begin{equation}
M_{mn}(\kk)=
\sum_{\GG}
\ip{\psi\bmk}{\kk+\GG} \ip{\kk+\GG}{\pt\bnk} \,,
\label{equa:plane-wave}
\end{equation}
where $\ket{\kk+\GG}$
is the plane wave $e^{i(\kk+\GG)\cdot\rr}$ for reciprocal vector $\GG$
normalized to the unit cell.
The evaluation should also be straightforward in other first-principles
basis sets.
For simple lattice models the Hamiltonian is typically
written in an orthonormal tight-binding basis, so that the
wavefunctions are
\begin{equation}
\ket{\psi\bnk}=\sum_j C_{nj,\kk}\,\ket{\chi_{j\kk}}
\label{equa:psichi}
\end{equation}
where $\ket{\chi_{j\kk}}$ are the Bloch basis states
\begin{equation}
\chi_{j\kk}(\rr)=\sum_{\RR} e^{i\kk\cdot\RR} \,\varphi_j(\rr-\RR)
\label{equa:chi}
\end{equation}
and $\varphi_j(\rr-\RR)$ is the $j$'th tight-binding basis orbital
in unit cell $\RR$.  Then the spillage is trivially computed using
\begin{equation}
M_{mn}(\kk)=\sum_j C^*_{mj,\kk} \Ct_{nj,\kk} \,.
\label{equa:tboverlap}
\end{equation}

Since the use of Wannier interpolation methods
\cite{MLWF-2,yates-prb07,MLWF-rmp}
is becoming increasingly frequent, we also comment on this case
here.  In this approach, the occupied Bloch states are again written
as in Eq.~(\ref{equa:psichi}), but this time the Bloch basis states are
\begin{equation}
\chi_{j\kk}(\rr)=\sum_{\RR} e^{i\kk\cdot\RR} \,w_j(\rr-\RR)
\label{equa:chiw}
\end{equation}
where $w_j(\rr-\RR)$ is the $j$'th Wannier function in unit cell $\RR$.
Then the spillage is again computed using Eqs.~(\ref{equa:def2}) and
(\ref{equa:tboverlap}).
This will be accurate as long as the WFs for the systems with and
without SOC are chosen to be the same, or as similar as possible.
As we shall see in the following section, the results from the
Wannier basis match those of the direct plane-wave calculation
very closely for the cases studied here.

In the case of complex unit cells or supercells with
many bands near the gap, it may be difficult to
identify precisely which bands have been inverted by the SOC.
In this case it may be helpful to define a
valence-band-resolved spillage
as
%
$
\gamma_{n}(\kk)=[L(\kk)L^{\dagger}(\kk)]_{nn},
$
%
where $L_{nm}(\kk)\!=\!\ip{\psi\bnk}{\pt\bmk}$ is the overlap
matrix between the occupied states without SOC and the unoccupied states
with SOC. Then the total spillage is
$\gamma(\kk)\!=\!\sum_{n=1}^{n\bocc}\gamma_{n}(\kk)$.
Similarly, $\bar{\gamma}_{m}=(L^{\dagger}L)_{mm}$
provides a conduction-band-resolved spillage. However,
it should be noted that $\gamma_n$ and $\bar{\gamma}_m$
are not gauge-invariant;
they will change under a unitary transformation among the occupied
or unoccupied states.
A natural gauge choice is the one associated with the singular-value
decomposition $L\!=\!V\Sigma W^{\dagger}$.  Transforming the sets
of occupied and unoccupied states according to the unitary matrices
$V$ and $W$ respectively, the overlap matrix between the transformed
states is just $\Sigma$, which is real and diagonal. The columns of
$V$ ($W$) corresponding to the leading eigenvalues indicate which
linear combinations of valence (conduction) states contribute the most
to the total spillage. We leave the exploration of these refinements
for a future study.

\subsection{Relation to topological character}
\label{sec:topo}

Here we argue that the presence of non-trivial topological
indices will be reflected in certain features of the spillage
distribution in the BZ.

We first consider the relatively simple
case in which the SOC-driven band inversion is associated with
the crossing of highest valence and lowest conduction states
belonging to two different irreducible representations (irreps)
at a high-symmetry point $\kk\!=\!\Lambda_0$ in the BZ.
Since the states belonging to different irreps have no overlap with
each other,
the spillage at $\Lambda_0$ must be greater than or equal
to the irrep dimension.  In TR-invariant \bise, for example,
the four states around the Fermi level at $\Gamma$ consist of two
Kramers doublets of opposite parity.  In this case the dimension of the
irreps is two, so we expect a peak in $\gamma(\kk)$ centered at
$\Gamma$ whose height is $\gamma_{\rm max}\!\ge\!2$.
As we shall show in Sec.~\ref{sec:materials}, this is exactly what we find
in \bise.

Next, we argue that a correspondence between topological
character and spillage should also remain valid for more
general cases without special lattice symmetry.  Let us first
consider the case of CIs (i.e., with broken TR symmetry).
We assume the Bloch functions $\psi\bnk$ are those of a normal
system with Chern number $C\!=\!0$, while $\pt\bnk$ are topologically
nontrivial with a nonzero Chern number $\Ct$.  We argue that this
implies the existence of at least one point in the BZ where the
spillage is $\ge1$.  If we assume the contrary, i.e., $\gamma(\kk)<1$
everywhere in the BZ, then the determinant of the overlap matrix
of Eq.~(\ref{equa:Mmn}) between $\psi\bnk$ and $\pt\bnk$ obeys
$\det(M_\kk)>0$ everywhere, since
a singular $M$ would imply $\gamma\ge1$.  Because the system
$\ket{\psi\bnk}$ is topologically normal, we know it is possible
to choose a smooth and periodic gauge for it, and we assume without
loss of generality that this has been done.  But if $M_\kk$ is
nowhere singular, the $\ket{\psi\bnk}$ can be used as ``trial
functions'' to construct a smooth and periodic gauge for the
$\ket{\pt\bnk}$, as follows.  At each
$\kk$, carry out a singular value decomposition to express
$M=V^\dagger\Sigma W$ ($V$ and $W$ are unitary and $\Sigma$
is real positive diagonal), and then use the unitary matrix $V^\dagger W$
to transform the original $\pt\bnk$ to a new set $\pt'\bnk$.  Then
$M=V^\dagger\Sigma V$, i.e., it is Hermitian and positive definite.
Intuitively, this means that a smooth and periodic
gauge has been chosen for the
states $\pt'\bnk$ to make them ``maximally aligned'' with the
states $\psi\bnk$.  But a smooth and periodic gauge is inconsistent
with a nonzero Chern number, completing the proof by contradiction.
Thus, if $\gamma<1$ everywhere in the BZ, then $M_{mn,\kk}$ is
nonsingular everywhere, and the system $\ket{\pt\bnk}$ is normal.
Conversely, a topological system must have $\gamma(\kk)\ge1$
somewhere in the BZ, which provides both a signal for the topological
phase and an indication of where in the BZ the band inversion has
occurred.

For the TR-invariant \Z2\ TIs, similar arguments can be put forward
that work even in the absence of inversion symmetry. If the system
of $\ket{\psi\bnk}$ is in the \Z2-even\ phase, one can always make
a smooth gauge choice over the entire BZ that respects TR symmetry.
In the \Z2-odd\ case, however, such a gauge choice does not
exist.\cite{fu-prb06,soluyanov-prb12} Therefore, $\det(M_\kk)$ must
vanish somewhere in BZ, or else the smooth gauge could be transferred
to the $\ket{\pt\bnk}$, resulting in a contradiction.  Due to the
TR symmetry, $\det(M_\kk)=\det(M_{-\kk})$, so one would generically expect
$\gamma(\kk)\!\geq\!1$ at two points ($\kk_0$ and $-\kk_0$) in the
BZ.  For the case of inversion-symmetric TIs, $\kk_0$ and $-\kk_0$
merge at one of the TRIM, the two spillages add up, and one expects
$\gamma\ge2$ at one of the TRIM.

In the following section, we numerically test and confirm
the above arguments by applying the formalism to systems in different
topological phases.

\section{Applications}

\subsection{Application to two-band Dirac Hamiltonian}
\label{sec:dirac}

As a warm-up exercise, we first apply the spillage formula
to a minimal model of a band inversion in 2D $(k_x,k_y)$ space,
namely a Dirac model at half filling as described by the Hamiltonian
\begin{align}
H\!=\!m(1-\lambda)\sigma_z+k_x\sigma_x+k_y\sigma_y
\end{align}
where $\sigma_j$ are Pauli matrices.  Here $m$ is a mass and
$\lambda$ is a control parameter that inverts the bands at
$\lambda\!=\!1$.
Physically, such a model may describe the low-energy physics
in the vicinity of a band touching event associated with the
transition from a normal to a quantum anomalous Hall insulator,
or at one of the band touching events (at $\kk_0$ or $-\kk_0$)
in the transition to a spin-Hall insulator.
The energy spectrum of the above Hamiltonian is
$E_{\pm}\!=\!\pm\sqrt{m^2(1-\lambda)^2+k_x^2+k_y^2}$, where
the gap closes at $\lambda\!=\!1$ at $\Gamma$ ($k_x\!=\!k_y\!=\!0$).
The spillage is just
$\gamma(\lambda,\kk)\!=\!1-\vert\ip{\psi_{1\kk}^{0}}{\psi_{1\kk}^{\lambda}}\vert^2$,
where $\ket{\psi_{1\kk}^{0}}$ ($\ket{\psi_{1\kk}^{\lambda}}$) is the occupied
eigenstate at zero (non-zero) $\lambda$.

\begin{figure}
\centering
\includegraphics[width=7cm]{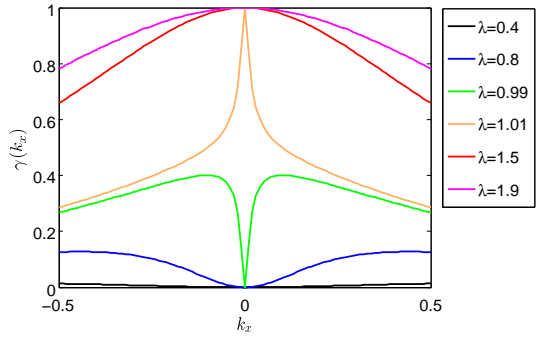}
\caption{The spillage of the half-filled Dirac
Hamiltonian as $\lambda$ increases from 0.4 to 1.9. }
\label{fig:dirac}
\end{figure}

Figure \ref{fig:dirac}
shows the spillage \textit{vs.} $k_x$ at $k_y\!=\!0$
as $\lambda$ is increased
from 0.4 to 1.9. When $\lambda\!=\!0.4$, the spillage is negligible
almost everywhere, and is exactly zero at $\Gamma$.  On the other
hand, when $\lambda\!=\!0.99$, which is very close
to the gap closure point, one finds two peaks of spillage
emerging on either side of $\Gamma$, with a peak value approaching
0.5 as $\lambda\rightarrow 1$. As $\lambda$  passes
through the critical point at $\lambda\!=\!1$, one finds that
the spillage at $\Gamma$ jumps from 0 to 1, and then gradually
spreads out in BZ as $\lambda$ is increased further.

This interesting behavior can be interpreted as follows.
When $\lambda\!=\!0$, the $\sigma_z$ term dominates around
$\Gamma$, so that the pesudospin is mostly along
the $z$ direction around $\Gamma$.
On the other hand, if $\lambda$ is very close to 1,
the $\sigma_x$ and $\sigma_y$ terms dominate near
(but not exactly at) $\Gamma$, forcing
the pseudospin direction to point in the $(x, y)$ plane and
resulting in a spillage of 1/2.
However, the $\sigma_x$ and $\sigma_y$ terms
vanish at $\Gamma$, which means the pseudospin has
to point along the $\pm z$ direction.  Therefore,
when $\lambda\!<\!1$ ($\lambda\!>\!1$),
the pseudospin is parallel (anti-parallel) with the pseudospin direction
at $\lambda\!=\!0$, such that the spillage jumps from 0 to 1 as $\lambda$
passes through the critical point.

\subsection{Application to the Kane-Mele model}
\label{sec:kane-mele}

The Kane-Mele model is a four-band TB model on a graphene lattice,
including nearest-neighbor (NN) spin-independent
hoppings and both NN and next-NN spin-dependent hoppings:
\begin{align}
H\!=\!& \sum_{\langle ij\rangle} t c^{\dagger}_{i} c_{j}+
\sum_{\langle\!\langle ij\rangle\!\rangle}i
\lambda\bso\nu_{ij}c^{\dagger}_{i}s_z c_{j}\notag\\
& +\sum_{\langle ij\rangle}i\lambda\bR c^{\dagger}_{i}
(\mathbf{s}\times\hat{\mathbf{d}}_{ij})_{z}c_{j}
+\sum_{i}{\epsilon(-1)^{i}c^{\dagger}_{i}c_{i}} \,.
\label{equa:kane-mele}
\end{align}
Here spin is implicit,
$t$ is the NN spin-independent hopping amplitude, $\lambda\bso$ is
the strength of the next-NN non-spin-flip SOC,
$\lambda\bR$ is the NN Rashba-like SOC amplitude, and $\epsilon$ is the
magnitude of on-site energy, with signs $\pm 1$ for A and B
sublattices respectively.  Also, $\nu_{ij}\!=\!\pm 1$ with the
sign depending on the chirality of the
next-NN bond from site $i$ to $j$, and $\hat{\mathbf{{d}}}_{ij}$ is
the unit vector pointing from site $i$
to its NN $j$.
In this model, $\lambda\bso$
competes with $\lambda\bR$ and $\epsilon$, in the sense that
$\lambda\bso$ tends to drive the system to the QSH phase
while $\lambda\bR$ and $\epsilon$ tend to retain the trivial band topology.

For simplicity, we first drop the Rashba coupling, so that
spin is a good quantum number.  The system
is in the QSH phase when $3\sqrt{3}\lambda\bso\!>\!\epsilon$,
and in the normal phase otherwise.  Without the Rashba term,
the Kane-Mele model can be considered as a superposition of two
copies of the Haldane model with opposite Chern numbers.\cite{Haldane-model}
If one calculates the 2D Chern numbers for spin-up and spin-down electrons
separately, one would find that the two Chern
numbers are $\pm1$ in the QSH phase.  While the Haldane-model system
goes from a normal insulator to a CI via a band inversion at
either the K or K$'$ point, the Kane-Mele model transitions to the
QSH state via simultaneous band inversions at both K and K$'$, 
but for opposite spins at these two points.

\begin{figure}
\centering
\subfigure{
\includegraphics[width=7.0cm,bb=18 251 559 552]{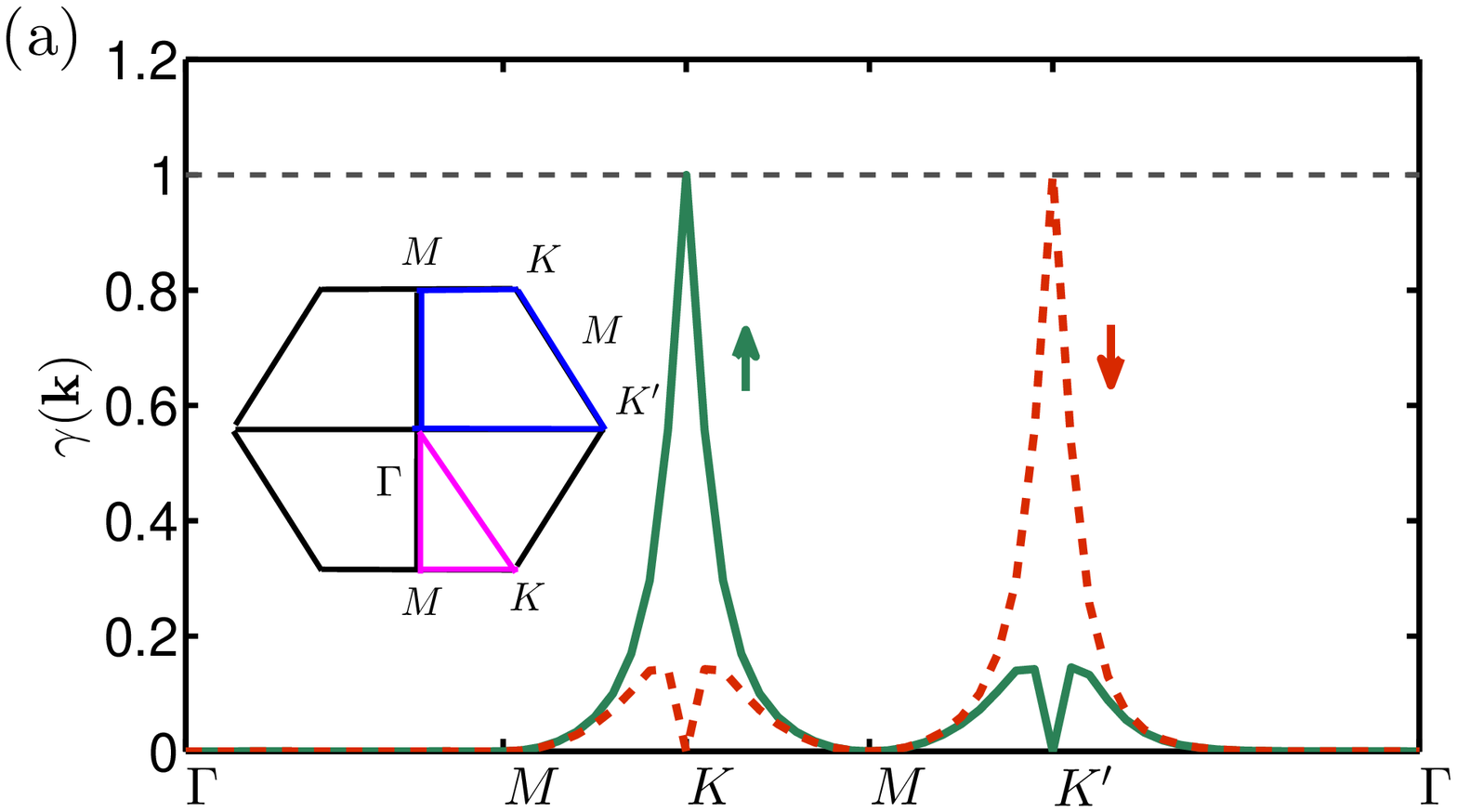}}
\subfigure{
\includegraphics[width=7.0cm,bb=44 256 536 544]{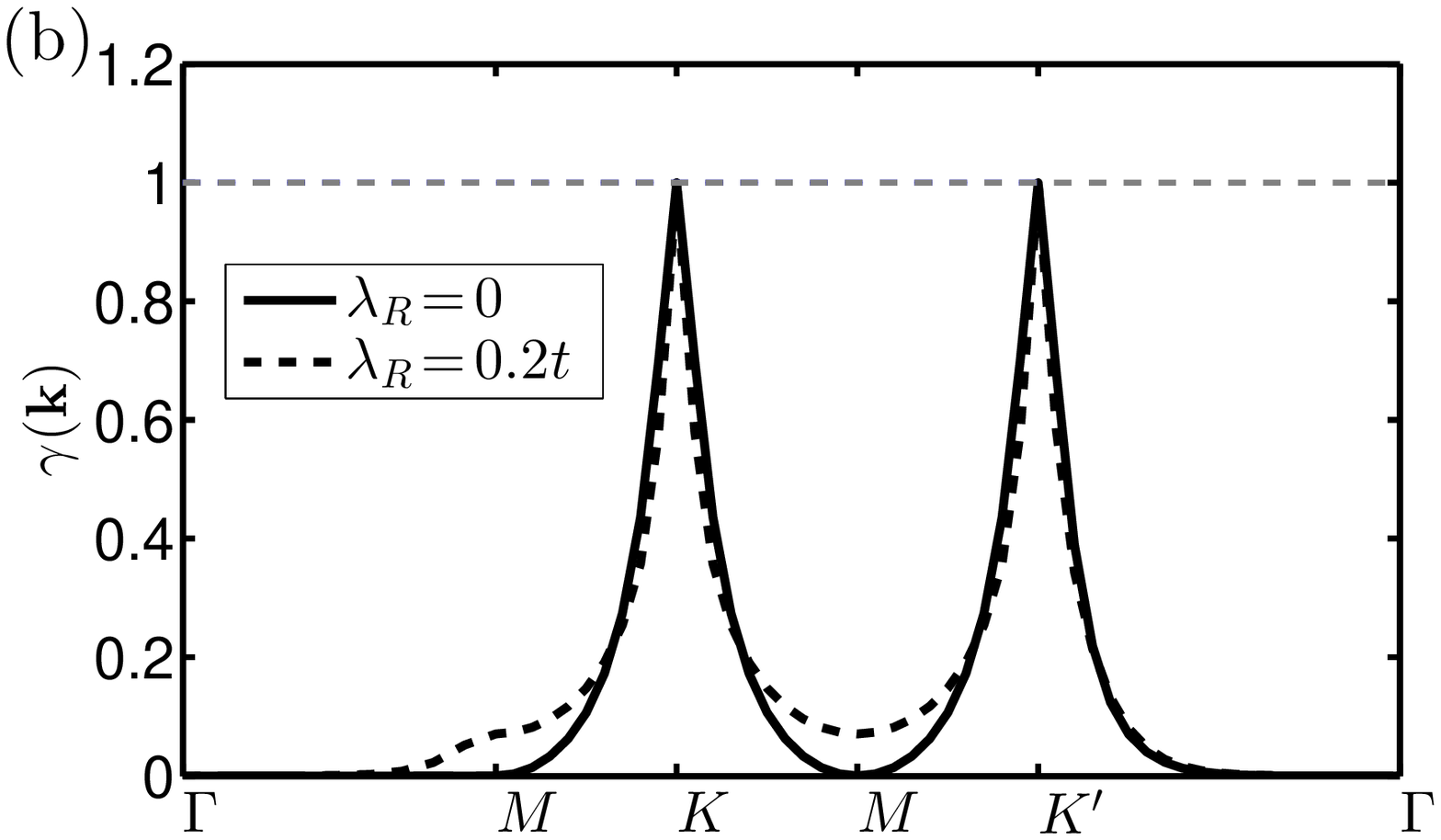}}
\caption{Spin-orbit spillage of the Kane-Mele model in the QSH phase,
with $t\!=\!1$, $\lambda\bso\!=\!0.1t$, and $\epsilon\!=\!0.1t$.
(a) Spin-resolved spillage without Rashba coupling; solid (green) and
dashed (red) lines denote spin-up and spin-down spillage.
Inset shows
$\Gamma$-M-K-M-K$^\prime$-$\Gamma$ path used here (blue) and
K-$\Gamma$-M-K path used in Fig.~\ref{fig:bi}
(magenta).
(b) Total spillage without (solid line) and with
(dashed line) Rashba coupling.}
\label{fig:km}
\end{figure}

The SOC-induced spillage without the Rashba term is
shown in Fig.~\ref{fig:km}(a).  In this case the spins act
independently, so the spin-up and spin-down spillages
$\gamma_\sigma(\mathbf{k})= n\bocc/2-\sum_{m,n=1}^{n\bocc/2}
\vert M_{n\sigma,m\sigma}(\mathbf{k})\vert^2$
(where $\sigma\!=\!\{\uparrow,\downarrow\}$)
are shown separately.  Clearly the spin-up band inversion at K is
responsible for $\gamma_{\uparrow}\!=\!1$, and conversely at
K$'$.  The total spillage
$\gamma(\mathbf{k})=\gamma_{\uparrow}(\mathbf{k})+
\gamma_{\downarrow}(\mathbf{k})$
is shown by the solid line in Fig.~\ref{fig:km}(b).  The symmetry between
the behavior at K and K$'$ has been restored by summing over spins.
Note that the peak values are $\gamma=\!1$ exactly; the fact that they
do not exceed one is an artifact of the simplicity of the model.
It is also interesting to note that in the absence of time-reversal symmetry,
the spin-resolved spillage is closely related to the van Vleck paramagnetism 
in spin-orbit coupled systems.

When the Rashba coupling is included, as shown by the dashed line
in Fig.~\ref{fig:km}(b), spin is no longer a good quantum number,
so that a spin decomposition is not well-defined.
As expected, adding the Rashba term does not significantly change the
results;\cite{comment_rashba}
one still finds that the spillage reaches unity at
K and K$'$ as before, providing an indication of the spin-Hall phase.

\subsection{Application to Chern insulators}
\label{sec:ci}

We now consider the case of broken TR symmetry, so that the
\Z2\ index is no longer well-defined, but the possibility of
CI phases appears.  As discussed in Sec.~\ref{sec:intro}, 
SOC is important here as well.  Here we study a buckled honeycomb
Bi bilayer with a Zeeman field applied normal to the plane, 
which can be regarded as having been cut from a
3D Bi crystal on a $(111)$ plane.
The Bi $(111)$ bilayer has been proposed as 
a candidate for QSH insulator.\cite{murakami-prl06}
If a Zeeman field is further applied,
it is possible to obtain CI phases with Chern numbers
$C\!=\!1$ or $C\!=\!-2$.\cite{hongbin-prb12,hongbin-prb13}
To describe this system we use a TB model based on Bi 6$s$ and
6$p$ orbitals,
where the first-neighbor $ss$,
$sp$, $pp\sigma$, and $pp\pi$ hoppings, as well as
the second-neighbor $pp\sigma$ hoppings, are included.
The hopping parameters
are taken from a TB model for 3D bulk Bi.\cite{bi-tb}
In order to obtain non-zero Chern numbers,
an on-site $p$-shell SOC ($\lambda_{\rm SOC}$)
and a Zeeman field ($H_z$) are further applied.
It turns out that if $H_z$ is fixed at 0.8\,eV,
then the phases with $C\!=\!-2$ and $+1$ are realized at
$\lambda_{\rm SOC}\!=\!2.4$\,eV and 0.6\,eV respectively.
If the SOC is completely turned off, $C\!=\!0$.

\begin{figure}
\centering
\subfigure{\includegraphics[height=4.0cm,bb=50 256 539 540]{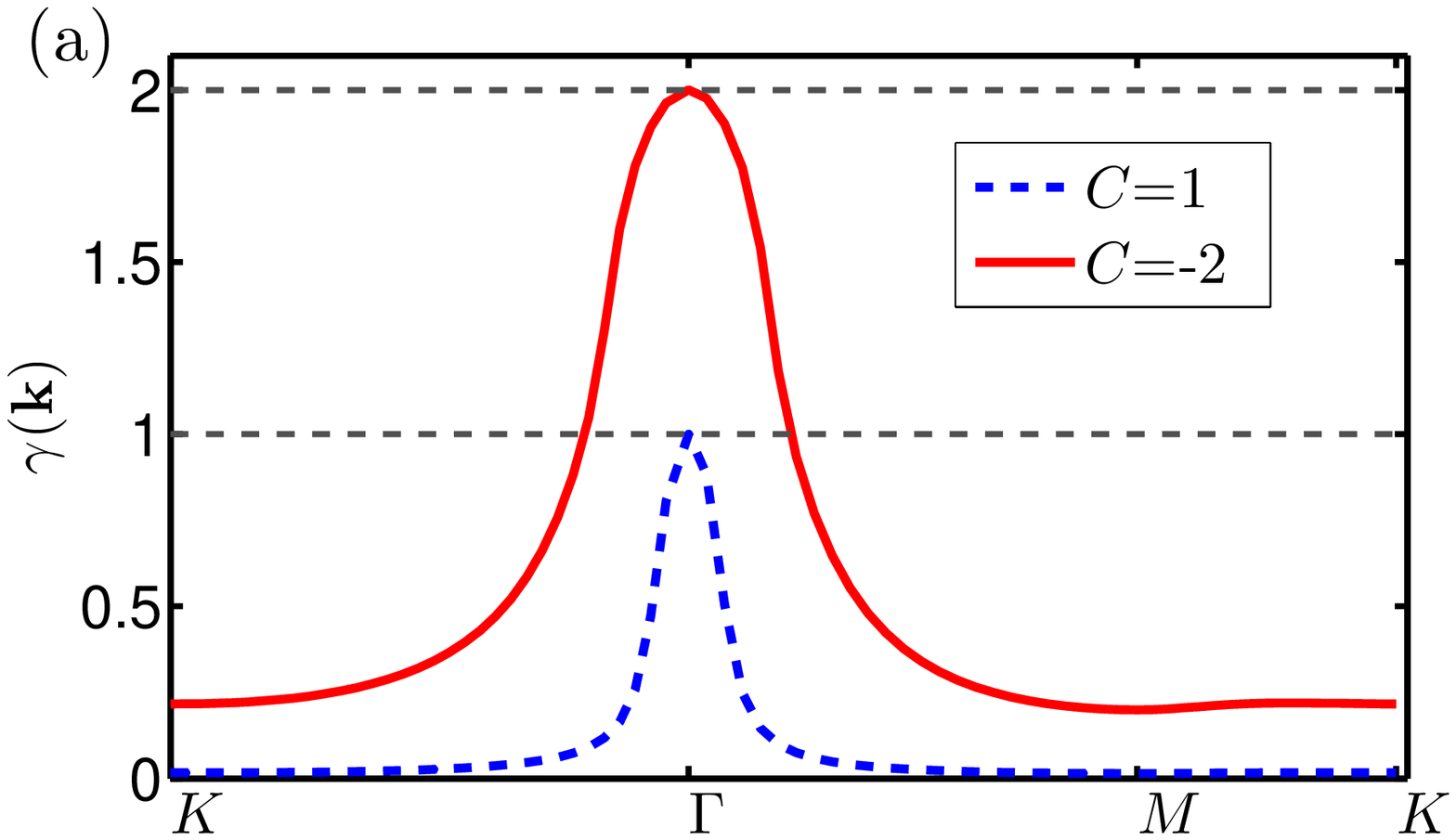}}
\subfigure{\includegraphics[height=4.2cm,bb=1 233 625 566]{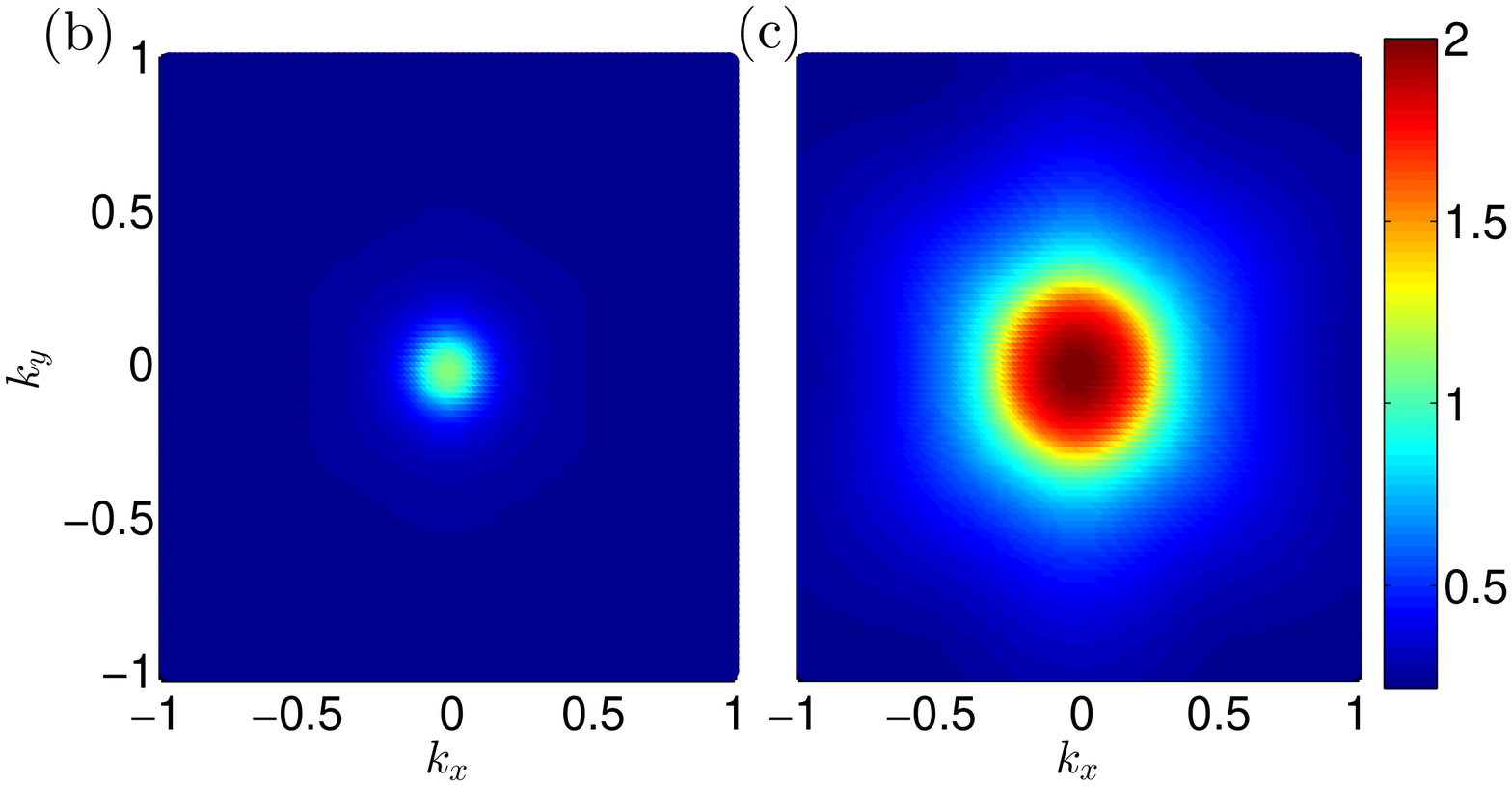}
}
\caption{(a) Spin-orbit spillage of the Bi bilayer
for $C\!=\!1$ (dashed blue) and
$C\!=\!-2$ (solid red) phases, plotted along the K-$\Gamma$-M-K path
(magenta path in inset of Fig.~\ref{fig:km}(a)).
(b) Spillage for $C\!=\!1$ phase plotted in the 2D BZ
($k_x$ and $k_y$ in units of \angstrom$^{-1}$).
(c) Same for $C\!=\!-2$ phase.}
\label{fig:bi}
\end{figure}

The spillage for the Bi bilayer is shown along a high-symmetry
$k$-path in Fig.~\ref{fig:bi}(a), and as a distribution in the 2D BZ in
Figs.~\ref{fig:bi}(b-c), for the two parameter sets giving the
$C\!=\!1$ and $C\!=\!-2$ phases.
In both cases the spillage distribution is concentrated at $\Gamma$,
indicating a band inversion there, although it is much more
sharply peaked in the $C\!=\!1$ case.
Clearly the spillages provide a signature of the presence of
a Chern-insulator phase, including the location of the band inversion and
the magnitude (but not the sign) of the Chern number.  Here again
the peak values of the spillage are exactly equal to the magnitude
of the Chern number.  For more realistic systems with more bands 
included, the spillage can be expected to exceed these values slightly,
but a clear correlation between the peak values of spillage and the
Chern number is still expected.

\subsection{Application to 3D topological insulators}
\label{sec:materials}

%
\begin{figure}
\centering
\includegraphics[width=6cm]{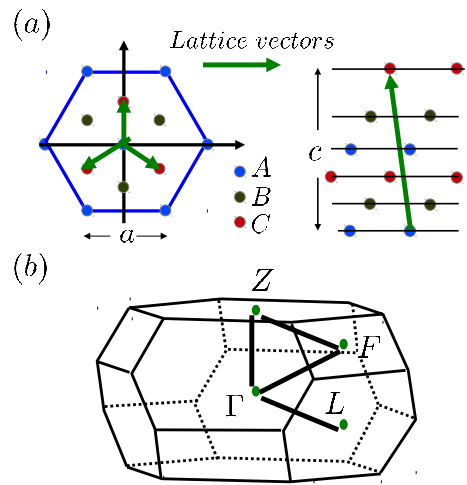}
\caption{(a) Lattice structure of \bise.
(b) The BZ of \bise; the spillage and bandstructures shown in
Fig.~\ref{fig:materials}(a) and Fig.~\ref{fig:proj} are plotted along the 
black path. }
\label{fig:lattice}
\end{figure}
In this subsection we apply our formalism to realistic first-principles
calculations of \bise, \inse\ and \sbse. \bise\ is a
well-known strong TI,\cite{zhang-np09} where the SOC-induced band inversion
takes place at $\Gamma$. We also consider \inse\ and \sbse\ in the same
crystal structure (known as $\beta$ phase for \inse\ and not realized
experimentally for \sbse), which are theoretically predicted
(and experimentally confirmed for \inse) to be trivial insulators.
\cite{zhang-np09,oh-prl12,armitage1,jpl-prb13}
Here it is interesting to note that even though Sb and
In have very similar atomic SOC strength, the substitution of In atoms
tends to drive \bise into a trivial-insulator phase
much faster than does Sb substitution, due to the existence of
In $5s$ orbitals.\cite{jpl-prb13}

As shown in Fig.~\ref{fig:lattice},
the considered structure is rhombohedral, with
two cations and three Se atoms in the primitive unit cell.
The five 2D monolayers are stacked in an
\hbox{$A$-$B$-$C$-$A$-...} sequence
along the (111) direction to form quintuple layers (QLs).  Experimentally
the in-plane hexagonal parameters are $a\!=\!4.138$ and 4.05\,\AA,
and the QL size is $c\!=\!9.547$ and 9.803\,\AA, for \bise\ and
\inse\ respectively.   In our calculations, we take the experimental 
lattice parameters of \bise\ and \inse, but relax their internal atomic
coordinates.  As for \sbse, because its rhombohedral structure 
is not adopted in nature, both the lattice parameters and atomic positions
are relaxed.
The ground state of rhombohedral \sbse\ is predicted to be a
trivial insulator with $a\!=\!4.11$\,\AA\ and $c\!=\!10.43$\,\AA.

We use the \textsc{Quantum ESPRESSO} package\cite{QE-2009} to carry out
first-principles calculations on these systems both with and without SOC.
The PBE generalized-gradient
approximation (GGA) is taken to treat the exchange-correlation functional,
\cite{pbe-1,pbe-2} and norm-conserving
pseudopotentials are generated from OPIUM package.\cite{opium-web,opium-paper}
The energy cutoff is taken as 65\,Ry for \inse\ and 55\,Ry for
\bise\ and \sbse, with an
8$\times$8$\times$8 Monkhorst-Pack $\bf k$ mesh.\cite{monkhorst-pack}
The wavefunctions defined in the plane-wave basis are extracted from these
calculations and Eq.~(~\ref{equa:plane-wave}) is applied to evaluate
the spillage.

As mentioned in Sec.~\ref{sec:intro}, the spillage can also be calculated
in the Wannier basis. Starting from the first-principles calculations,
we use the \textsc{Wannier90} package\cite{wannier90} to
construct Wannier functions (WFs) and a
corresponding realistic TB model\cite{comment_wannier90} for
each of the three materials.
The basis WFs are constructed by projecting 30 atomic $p$ trial
orbitals onto the Bloch subspace of $p$-like bands to generate a
30-band spinor model for \bise\ and \sbse, 
whereas four additional In $5s$ projectors and bands are
included in the model for \inse.
In order that they will retain their atomic-like identity as much
as possible, the projected WFs are not optimized to minimize the
spread functional.\cite{MLWF-rmp} We find that the
WFs generated by this projection method are almost the same for the 
systems with and without SOC,
so that the matrix elements $M_{mn}(\kk)$ defined in Sec.~\ref{sec:def}
can be evaluated with good accuracy.
\begin{figure}
\centering
\subfigure{
\includegraphics[height=4.7cm,bb=34 236 551 556]
{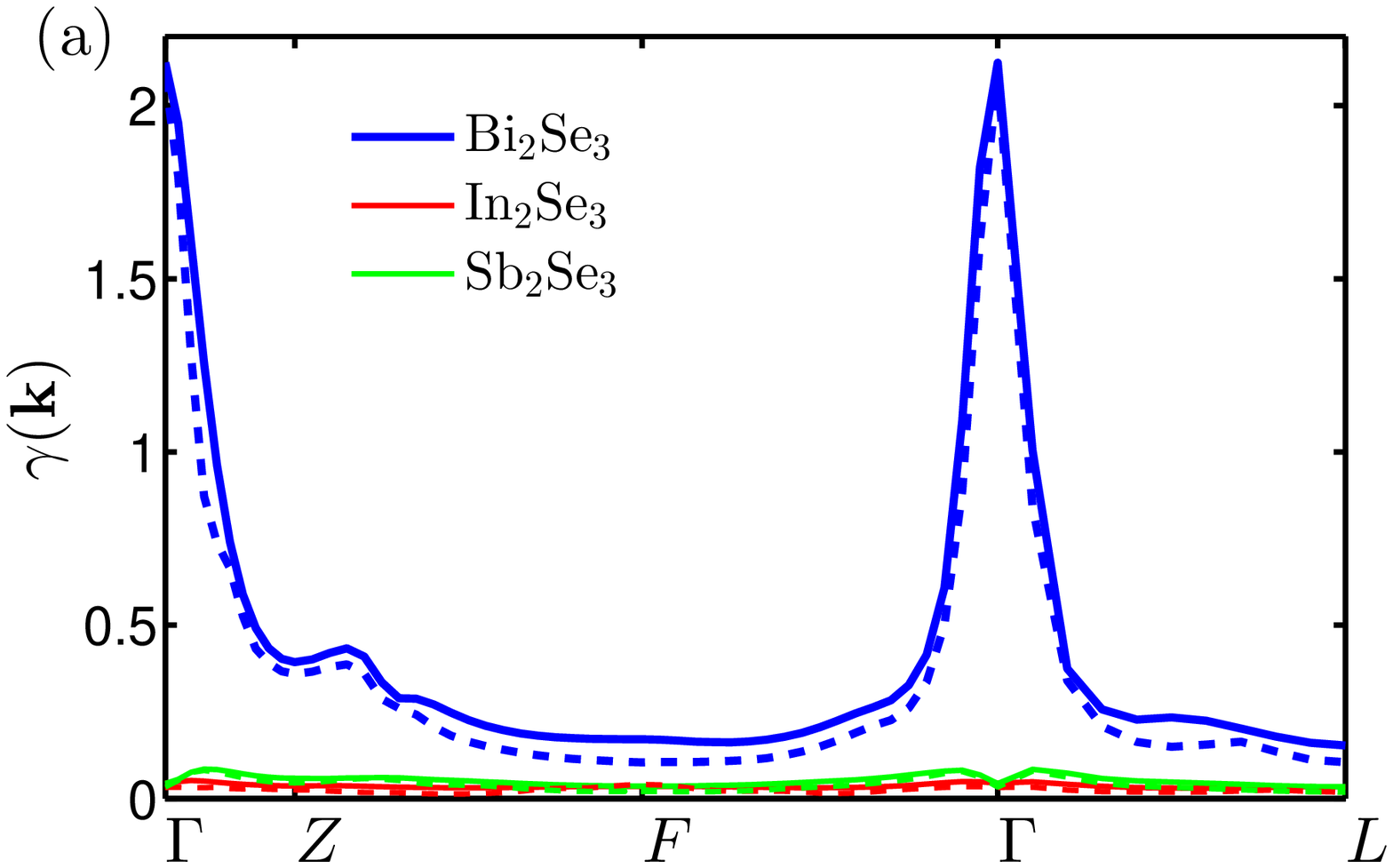}}
\subfigure{\includegraphics[width=7.0cm,bb=86 246 515 545]{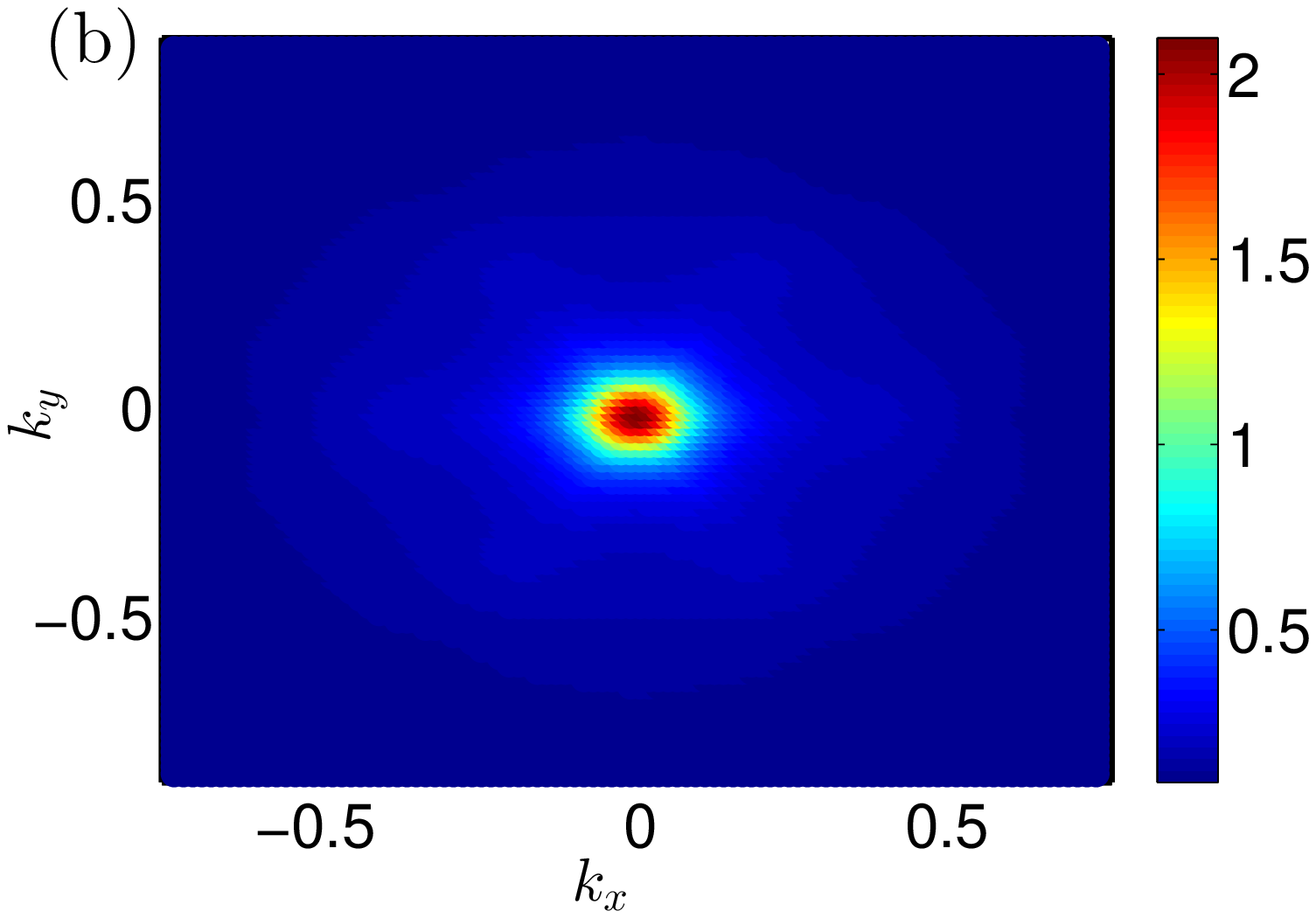}
}
\caption{(a) Spin-orbit spillage of rhombohedral 
\bise, \sbse\ and \inse\ as indicated by blue, green and red lines
respectively.  Solid (dashed) lines show the spillage computed
from direct plane-wave (Wannier-based) calculations.
(b) Spillage of \bise\ in the
$(k_x,k_y)$ plane at $k_z\!=\!0$
(units of \angstrom$^{-1}$).}
\label{fig:materials}
\end{figure}

The spillage from the direct plane-wave calculations are shown as
the solid lines in Fig.~\ref{fig:materials}(a).
For \bise, the spillage $\gamma(\mathbf{k})$ has a peak value of 2.12
at $\Gamma$, which is slightly larger than 2, indicating that two
Kramers degenerate bands at $\Gamma$ have been inverted by SOC.
On the other hand, the effect of SOC in \inse\ and \sbse\ seems to be
negligible everywhere in the BZ, which is consistent with the fact
that they are both trivial insulators.

\begin{figure}
\centering
\includegraphics[width=8.5cm,bb=33 252 579 555]{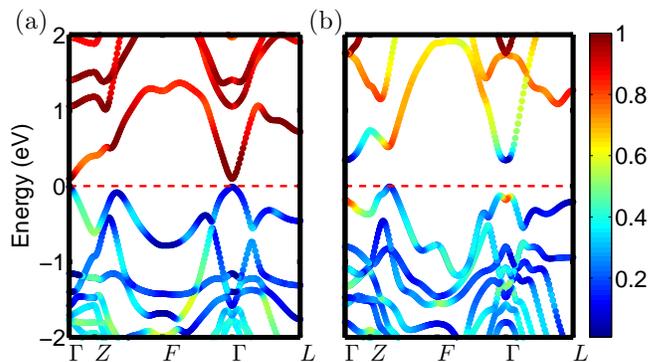}
\caption{
(a) Wannier-interpolated bandstructure of \sbse.
(b) Same for \bise. Color coding indicates weight of
Sb $5p$ or Bi $6p$ orbitals. }
\label{fig:proj}
\end{figure}

The calculations carried out in the Wannier basis are shown by
the dashed lines in Fig.~\ref{fig:materials}(a).  The spillage
is typically slightly larger for the direct plane-wave
calculations, since the fact that the WFs have a slightly
different plane-wave representation with and without SOC is not
taken into account in the Wannier-based calculations.
Still, the qualitative features are the same,
showing that the Wannier-based approach can successfully provide the
same kind of information about the nature and location of the
topological band inversion. In
Fig.~\ref{fig:materials}(b) we also show the spillage
of \bise\ in the ($k_x, k_y$) plane at $k_z\!=\!0$, as calculated
in the Wannier basis,
which again indicates a highly localized band inversion near
$\Gamma$ and is fully consistent with the expected picture of
the band inversion in \bise.

To see the band inversion from another perspective, we plot
in Fig.~\ref{fig:proj}
the bulk bandstructures of \sbse\ and \bise\ projected onto
Sb $5p$ and Bi $6p$ orbitals respectively.  It is clear that
for \sbse, the Sb $5p$ orbitals are almost all concentrated in the
conduction bands, whereas in \bise\ there is a localized region
around $\Gamma$ where the corresponding Bi $6p$ orbitals
contribute mostly to the top valence band.  This is precisely
the region of the band inversion
corresponding to the peak at $\Gamma$ in Fig.~\ref{fig:materials}.

\section{Summary}

To summarize, we have introduced the SOC-induced spillage $\gamma(\kk)$
as a useful quantitative tool for evaluating the degree of band inversion
driven by SOC and mapping it as a function of $\kk$ in the BZ.
We have applied this method to the two-band Dirac model,
the 2D Kane-Mele model and a
tight-binding model of a Bi bilayer with applied Zeeman field,
as well as to realistic materials including
both trivial and topological insulators. A clear correspondence
between non-trivial topological indices and non-trivial spillage
distributions is evident. In the two-band Dirac model, one observes
interesting behavior in the distribution of spillage through a topological
phase transition process. In the Kane-Mele model,
one gets two peaks of spillage at K and K$'$ with the 
peak value of 1, which indicates that a single band is inverted
at these two points corresponding to an odd 2D \Z2\ index. 
In the Bi bilayer with applied Zeeman field,
a peak of spillage shows up at
$\Gamma$, with the peak value corresponding to the absolute value
of the Chern number. In \bise, the spillage is slightly
greater than 2 at one of the TRIM, namely $\Gamma$,
implying that two bands are inverted by SOC there and signaling
the presence of a nontrivial strong \Z2\ index.

As mentioned above, other methods exist for the direct computation
of topological Chern and \Z2\ indices, with or without inversion
symmetry,\cite{fukui-jpsj05,fu-prb07,soluyanov-prb11}
and we still recommend these if a direct and definitive determination
of the topological indices is needed.  However, the present spillage-based
approach has the advantage of providing a BZ map of the strength,
position, and degree of localization of the band inversion responsible
for the topological character, thus giving valuable physical intuition
about the origin of the topological properties of the material in question.
In addition, compared with
direct methods for topological index calculation,  the spillage
calculation only requires the evaluation of overlaps between two
wavefunctions at the same $\kk$ point, which is easy to implement
and numerically very efficient. Therefore, it is our hope that the
calculation of SOC spillage will prove to be a widely useful tool
that can be applied both for high-throughput screening for topological
materials and for obtaining a deeper understanding of the critical
features of the bandstructures in known topological materials.

\acknowledgments

This work was supported by NSF Grant DMR-10-05838.
We appreciate valuable discussions with Hongbin Zhang and Huaqing Huang.

\bibliography{spillage}

\begin{thebibliography}{10}%
\makeatletter
\providecommand \@ifxundefined [1]{%
 \ifx #1\undefined \expandafter \@firstoftwo
 \else \expandafter \@secondoftwo
\fi
}%
\providecommand \@ifnum [1]{%
 \ifnum #1\expandafter \@firstoftwo
 \else \expandafter \@secondoftwo
\fi
}%
\providecommand \enquote [1]{``#1''}%
\providecommand \bibnamefont  [1]{#1}%
\providecommand \bibfnamefont [1]{#1}%
\providecommand \citenamefont [1]{#1}%
\providecommand\href[0]{\@sanitize\@href}%
\providecommand\@href[1]{\endgroup\@@startlink{#1}\endgroup\@@href}%
\providecommand\@@href[1]{#1\@@endlink}%
\providecommand \@sanitize [0]{\begingroup\catcode`\&12\catcode`\#12\relax}%
\@ifxundefined \pdfoutput {\@firstoftwo}{%
 \@ifnum{\z@=\pdfoutput}{\@firstoftwo}{\@secondoftwo}%
}{%
 \providecommand\@@startlink[1]{\leavevmode\special{html:<a href="#1">}}%
 \providecommand\@@endlink[0]{\special{html:</a>}}%
}{%
 \providecommand\@@startlink[1]{%
  \leavevmode
  \pdfstartlink
   attr{/Border[0 0 1 ]/H/I/C[0 1 1]}%
   user{/Subtype/Link/A<</Type/Action/S/URI/URI(#1)>>}%
  \relax
 }%
 \providecommand\@@endlink[0]{\pdfendlink}%
}%
\providecommand \url  [0]{\begingroup\@sanitize \@url }%
\providecommand \@url [1]{\endgroup\@href {#1}{\urlprefix}}%
\providecommand \urlprefix [0]{URL }%
\providecommand \Eprint[0]{\href }%
\@ifxundefined \urlstyle {%
  \providecommand \doi [1]{doi:\discretionary{}{}{}#1}%
}{%
  \providecommand \doi [0]{doi:\discretionary{}{}{}\begingroup
  \urlstyle{rm}\Url }%
}%
\providecommand \doibase [0]{http://dx.doi.org/}%
\providecommand \Doi[1]{\href{\doibase#1}}%
\providecommand \bibAnnote [3]{%
  \BibitemShut{#1}%
  \begin{quotation}\noindent
    \textsc{Key:}\ #2\\\textsc{Annotation:}\ #3%
  \end{quotation}%
}%
\providecommand \bibAnnoteFile [2]{%
  \IfFileExists{#2}{\bibAnnote {#1} {#2} {\input{#2}}}{}%
}%
\providecommand \typeout [0]{\immediate \write \m@ne }%
\providecommand \selectlanguage [0]{\@gobble}%
\providecommand \bibinfo [0]{\@secondoftwo}%
\providecommand \bibfield [0]{\@secondoftwo}%
\providecommand \translation [1]{[#1]}%
\providecommand \BibitemOpen[0]{}%
\providecommand \bibitemStop [0]{}%
\providecommand \bibitemNoStop [0]{.\EOS\space}%
\providecommand \EOS [0]{\spacefactor3000\relax}%
\providecommand \BibitemShut [1]{\csname bibitem#1\endcsname}%
\bibitem{winkler2003}%
  \BibitemOpen
  \bibfield{author}{%
  \bibinfo {author} {\bibfnamefont{R.}~\bibnamefont{Winkler}},\ }%
  \emph{\bibinfo {title} {Spin-orbit coupling effects in two-dimensional
  electron and hole systems}},\ Vol.\ \bibinfo {volume} {191}\ (\bibinfo
  {publisher} {Springer Verlag},\ \bibinfo {year} {2003})%
  \bibAnnoteFile{NoStop}{winkler2003}%
\bibitem{nikolic2010oxford}%
  \BibitemOpen
  \bibfield{author}{%
  \bibinfo {author} {\bibfnamefont{B.}~\bibnamefont{Nikoli{\'c}}}, \bibinfo
  {author} {\bibfnamefont{L.}~\bibnamefont{Zarbo}},\ and\ \bibinfo {author}
  {\bibfnamefont{S.}~\bibnamefont{Souma}},\ }%
  \emph{\bibinfo {title} {The Oxford Handbook on Nanoscience and Technology:
  Frontiers and Advances}}\ (\bibinfo {publisher} {Oxford University Press},\
  \bibinfo {year} {2010})\ \bibinfo {note} {edited by A. V. Narlikar and Y. Y.
  Fu, Vol. I, Chap. 24, pp. 814–866}%
  \bibAnnoteFile{NoStop}{nikolic2010oxford}%
\bibitem{berry84}%
  \BibitemOpen
  \bibfield{author}{%
  \bibinfo {author} {\bibfnamefont{M.~V.}\ \bibnamefont{Berry}},\ }%
  \bibfield{journal}{%
  \bibinfo {journal} {Proceedings of the Royal Society of London. A.
  Mathematical and Physical Sciences}\ }%
  \textbf{\bibinfo {volume} {392}},\ \bibinfo {pages} {45} (\bibinfo {year}
  {1984})%
  \bibAnnoteFile{NoStop}{berry84}%
\bibitem{hln}%
  \BibitemOpen
  \bibfield{author}{%
  \bibinfo {author} {\bibfnamefont{S.}~\bibnamefont{Hikami}}, \bibinfo {author}
  {\bibfnamefont{A.~I.}\ \bibnamefont{Larkin}},\ and\ \bibinfo {author}
  {\bibfnamefont{Y.}~\bibnamefont{Nagaoka}},\ }%
  \bibfield{journal}{%
  \bibinfo {journal} {Progress of Theoretical Physics}\ }%
  \textbf{\bibinfo {volume} {63}},\ \bibinfo {pages} {707} (\bibinfo {year}
  {1980})%
  \bibAnnoteFile{NoStop}{hln}%
\bibitem{Hirsch99}%
  \BibitemOpen
  \bibfield{author}{%
  \bibinfo {author} {\bibfnamefont{J.~E.}\ \bibnamefont{Hirsch}},\ }%
  \bibfield{journal}{%
  \Doi{10.1103/PhysRevLett.83.1834}{\bibinfo {journal} {Phys. Rev. Lett.}}\ }%
  \textbf{\bibinfo {volume} {83}},\ \bibinfo {pages} {1834} (\bibinfo {month}
  {Aug}\ \bibinfo {year} {1999})%
  \bibAnnoteFile{NoStop}{Hirsch99}%
\bibitem{bhz06}%
  \BibitemOpen
  \bibfield{author}{%
  \bibinfo {author} {\bibfnamefont{B.~A.}\ \bibnamefont{Bernevig}}, \bibinfo
  {author} {\bibfnamefont{T.~L.}\ \bibnamefont{Hughes}},\ and\ \bibinfo
  {author} {\bibfnamefont{S.-C.}\ \bibnamefont{Zhang}},\ }%
  \bibfield{journal}{%
  \Doi{10.1126/science.1133734}{\bibinfo {journal} {Science}}\ }%
  \textbf{\bibinfo {volume} {314}},\ \bibinfo {pages} {1757} (\bibinfo {year}
  {2006})%
  \bibAnnoteFile{NoStop}{bhz06}%
\bibitem{konig06science}%
  \BibitemOpen
  \bibfield{author}{%
  \bibinfo {author} {\bibfnamefont{M.}~\bibnamefont{K{\"o}nig}}, \bibinfo
  {author} {\bibfnamefont{S.}~\bibnamefont{Wiedmann}}, \bibinfo {author}
  {\bibfnamefont{C.}~\bibnamefont{Br{\"u}ne}}, \bibinfo {author}
  {\bibfnamefont{A.}~\bibnamefont{Roth}}, \bibinfo {author}
  {\bibfnamefont{H.}~\bibnamefont{Buhmann}}, \bibinfo {author}
  {\bibfnamefont{L.~W.}\ \bibnamefont{Molenkamp}}, \bibinfo {author}
  {\bibfnamefont{X.-L.}\ \bibnamefont{Qi}},\ and\ \bibinfo {author}
  {\bibfnamefont{S.-C.}\ \bibnamefont{Zhang}},\ }%
  \bibfield{journal}{%
  \bibinfo {journal} {Science}\ }%
  \textbf{\bibinfo {volume} {318}},\ \bibinfo {pages} {766} (\bibinfo {year}
  {2007})%
  \bibAnnoteFile{NoStop}{konig06science}%
\bibitem{kane-prl05-a}%
  \BibitemOpen
  \bibfield{author}{%
  \bibinfo {author} {\bibfnamefont{C.~L.}\ \bibnamefont{Kane}}\ and\ \bibinfo
  {author} {\bibfnamefont{E.~J.}\ \bibnamefont{Mele}},\ }%
  \bibfield{journal}{%
  \bibinfo {journal} {Phys. Rev. Lett.}\ }%
  \textbf{\bibinfo {volume} {95}},\ \bibinfo {pages} {146802} (\bibinfo {year}
  {2005})%
  \bibAnnoteFile{NoStop}{kane-prl05-a}%
\bibitem{kane-prl05-b}%
  \BibitemOpen
  \bibfield{author}{%
  \bibinfo {author} {\bibfnamefont{C.~L.}\ \bibnamefont{Kane}}\ and\ \bibinfo
  {author} {\bibfnamefont{E.~J.}\ \bibnamefont{Mele}},\ }%
  \bibfield{journal}{%
  \bibinfo {journal} {Phys. Rev. Lett.}\ }%
  \textbf{\bibinfo {volume} {95}},\ \bibinfo {pages} {226801} (\bibinfo {year}
  {2005})%
  \bibAnnoteFile{NoStop}{kane-prl05-b}%
\bibitem{kane-rmp10}%
  \BibitemOpen
  \bibfield{author}{%
  \bibinfo {author} {\bibfnamefont{M.~Z.}\ \bibnamefont{Hasan}}\ and\ \bibinfo
  {author} {\bibfnamefont{C.~L.}\ \bibnamefont{Kane}},\ }%
  \bibfield{journal}{%
  \Doi{10.1103/RevModPhys.82.3045}{\bibinfo {journal} {Rev. Mod. Phys.}}\ }%
  \textbf{\bibinfo {volume} {82}},\ \bibinfo {pages} {3045} (\bibinfo {month}
  {Nov}\ \bibinfo {year} {2010})%
  \bibAnnoteFile{NoStop}{kane-rmp10}%
\bibitem{zhang-rmp11}%
  \BibitemOpen
  \bibfield{author}{%
  \bibinfo {author} {\bibfnamefont{X.-L.}\ \bibnamefont{Qi}}\ and\ \bibinfo
  {author} {\bibfnamefont{S.-C.}\ \bibnamefont{Zhang}},\ }%
  \bibfield{journal}{%
  \Doi{10.1103/RevModPhys.83.1057}{\bibinfo {journal} {Rev. Mod. Phys.}}\ }%
  \textbf{\bibinfo {volume} {83}},\ \bibinfo {pages} {1057} (\bibinfo {month}
  {Oct}\ \bibinfo {year} {2011})%
  \bibAnnoteFile{NoStop}{zhang-rmp11}%
\bibitem{yao-review12}%
  \BibitemOpen
  \bibfield{author}{%
  \bibinfo {author} {\bibfnamefont{W.}~\bibnamefont{Feng}}\ and\ \bibinfo
  {author} {\bibfnamefont{Y.}~\bibnamefont{Yao}},\ }%
  \bibfield{journal}{%
  \bibinfo {journal} {Science China Physics, Mechanics and Astronomy}\ }%
  \textbf{\bibinfo {volume} {55}},\ \bibinfo {pages} {2199} (\bibinfo {year}
  {2012})%
  \bibAnnoteFile{NoStop}{yao-review12}%
\bibitem{zhang-np09}%
  \BibitemOpen
  \bibfield{author}{%
  \bibinfo {author} {\bibfnamefont{H.}~\bibnamefont{Zhang}}, \bibinfo {author}
  {\bibfnamefont{C.-X.}\ \bibnamefont{Liu}}, \bibinfo {author}
  {\bibfnamefont{X.-L.}\ \bibnamefont{Qi}}, \bibinfo {author}
  {\bibfnamefont{X.}~\bibnamefont{Dai}}, \bibinfo {author}
  {\bibfnamefont{Z.}~\bibnamefont{Fang}},\ and\ \bibinfo {author}
  {\bibfnamefont{S.-C.}\ \bibnamefont{Zhang}},\ }%
  \bibfield{journal}{%
  \Doi{10.1038/nphys1270}{\bibinfo {journal} {Nature Physics}}\ }%
  \textbf{\bibinfo {volume} {5}},\ \bibinfo {pages} {438} (\bibinfo {year}
  {2009})%
  \bibAnnoteFile{NoStop}{zhang-np09}%
\bibitem{fu-prb07}%
  \BibitemOpen
  \bibfield{author}{%
  \bibinfo {author} {\bibfnamefont{L.}~\bibnamefont{Fu}}\ and\ \bibinfo
  {author} {\bibfnamefont{C.~L.}\ \bibnamefont{Kane}},\ }%
  \bibfield{journal}{%
  \bibinfo {journal} {Phys. Rev. B}\ }%
  \textbf{\bibinfo {volume} {76}},\ \bibinfo {pages} {045302} (\bibinfo {year}
  {2007})%
  \bibAnnoteFile{NoStop}{fu-prb07}%
\bibitem{Haldane-model}%
  \BibitemOpen
  \bibfield{author}{%
  \bibinfo {author} {\bibfnamefont{F.~D.~M.}\ \bibnamefont{Haldane}},\ }%
  \bibfield{journal}{%
  \Doi{10.1103/PhysRevLett.61.2015}{\bibinfo {journal} {Phys. Rev. Lett.}}\ }%
  \textbf{\bibinfo {volume} {61}},\ \bibinfo {pages} {2015} (\bibinfo {month}
  {Oct}\ \bibinfo {year} {1988})%
  \bibAnnoteFile{NoStop}{Haldane-model}%
\bibitem{hongbin-prb13}%
  \BibitemOpen
  \bibfield{author}{%
  \bibinfo {author} {\bibfnamefont{H.}~\bibnamefont{Zhang}}, \bibinfo {author}
  {\bibfnamefont{F.}~\bibnamefont{Freimuth}}, \bibinfo {author}
  {\bibfnamefont{G.}~\bibnamefont{Bihlmayer}}, \bibinfo {author}
  {\bibfnamefont{M.}~\bibnamefont{Le\ifmmode \check{z}\else
  \v{z}\fi{}ai\ifmmode~\acute{c}\else \'{c}\fi{}}}, \bibinfo {author}
  {\bibfnamefont{S.}~\bibnamefont{Bl\"ugel}},\ and\ \bibinfo {author}
  {\bibfnamefont{Y.}~\bibnamefont{Mokrousov}},\ }%
  \bibfield{journal}{%
  \Doi{10.1103/PhysRevB.87.205132}{\bibinfo {journal} {Phys. Rev. B}}\ }%
  \textbf{\bibinfo {volume} {87}},\ \bibinfo {pages} {205132} (\bibinfo {month}
  {May}\ \bibinfo {year} {2013})%
  \bibAnnoteFile{NoStop}{hongbin-prb13}%
\bibitem{klintenberg-archive10}%
  \BibitemOpen
  \bibfield{author}{%
  \bibinfo {author} {\bibfnamefont{M.}~\bibnamefont{Klintenberg}},\ }%
  \bibfield{journal}{%
  \bibinfo {journal} {arXiv:1007.4838}}%
   (\bibinfo {year} {2010})%
  \bibAnnoteFile{NoStop}{klintenberg-archive10}%
\bibitem{yang-nm12}%
  \BibitemOpen
  \bibfield{author}{%
  \bibinfo {author} {\bibfnamefont{K.}~\bibnamefont{Yang}}, \bibinfo {author}
  {\bibfnamefont{W.}~\bibnamefont{Setyawan}}, \bibinfo {author}
  {\bibfnamefont{S.}~\bibnamefont{Wang}}, \bibinfo {author}
  {\bibfnamefont{M.~B.}\ \bibnamefont{Nardelli}},\ and\ \bibinfo {author}
  {\bibfnamefont{S.}~\bibnamefont{Curtarolo}},\ }%
  \bibfield{journal}{%
  \bibinfo {journal} {Nature Materials}\ }%
  \textbf{\bibinfo {volume} {11}},\ \bibinfo {pages} {614} (\bibinfo {year}
  {2012})%
  \bibAnnoteFile{NoStop}{yang-nm12}%
\bibitem{dft1}%
  \BibitemOpen
  \bibfield{author}{%
  \bibinfo {author} {\bibfnamefont{P.}~\bibnamefont{Hohenberg}}\ and\ \bibinfo
  {author} {\bibfnamefont{W.}~\bibnamefont{Kohn}},\ }%
  \bibfield{journal}{%
  \Doi{10.1103/PhysRev.136.B864}{\bibinfo {journal} {Phys. Rev.}}\ }%
  \textbf{\bibinfo {volume} {136}},\ \bibinfo {pages} {B864} (\bibinfo {month}
  {Nov}\ \bibinfo {year} {1964})%
  \bibAnnoteFile{NoStop}{dft1}%
\bibitem{dft2}%
  \BibitemOpen
  \bibfield{author}{%
  \bibinfo {author} {\bibfnamefont{W.}~\bibnamefont{Kohn}}\ and\ \bibinfo
  {author} {\bibfnamefont{L.~J.}\ \bibnamefont{Sham}},\ }%
  \bibfield{journal}{%
  \Doi{10.1103/PhysRev.140.A1133}{\bibinfo {journal} {Phys. Rev.}}\ }%
  \textbf{\bibinfo {volume} {140}},\ \bibinfo {pages} {A1133} (\bibinfo {month}
  {Nov}\ \bibinfo {year} {1965})%
  \bibAnnoteFile{NoStop}{dft2}%
\bibitem{MLWF-2}%
  \BibitemOpen
  \bibfield{author}{%
  \bibinfo {author} {\bibfnamefont{I.}~\bibnamefont{Souza}}, \bibinfo {author}
  {\bibfnamefont{N.}~\bibnamefont{Marzari}},\ and\ \bibinfo {author}
  {\bibfnamefont{D.}~\bibnamefont{Vanderbilt}},\ }%
  \bibfield{journal}{%
  \bibinfo {journal} {Phys. Rev. B}\ }%
  \textbf{\bibinfo {volume} {65}},\ \bibinfo {pages} {035109} (\bibinfo {year}
  {2001})%
  \bibAnnoteFile{NoStop}{MLWF-2}%
\bibitem{yates-prb07}%
  \BibitemOpen
  \bibfield{author}{%
  \bibinfo {author} {\bibfnamefont{J.~R.}\ \bibnamefont{Yates}}, \bibinfo
  {author} {\bibfnamefont{X.~J.}\ \bibnamefont{Wang}}, \bibinfo {author}
  {\bibfnamefont{D.}~\bibnamefont{Vanderbilt}},\ and\ \bibinfo {author}
  {\bibfnamefont{I.}~\bibnamefont{Souza}},\ }%
  \bibfield{journal}{%
  \Doi{{10.1103/PhysRevB.75.195121}}{\bibinfo {journal} {Phys. Rev. B}}\ }%
  \textbf{\bibinfo {volume} {75}},\ \bibinfo {pages} {195121} (\bibinfo {year}
  {2007})%
  \bibAnnoteFile{NoStop}{yates-prb07}%
\bibitem{MLWF-rmp}%
  \BibitemOpen
  \bibfield{author}{%
  \bibinfo {author} {\bibfnamefont{N.}~\bibnamefont{Marzari}}, \bibinfo
  {author} {\bibfnamefont{A.~A.}\ \bibnamefont{Mostofi}}, \bibinfo {author}
  {\bibfnamefont{J.~R.}\ \bibnamefont{Yates}}, \bibinfo {author}
  {\bibfnamefont{I.}~\bibnamefont{Souza}},\ and\ \bibinfo {author}
  {\bibfnamefont{D.}~\bibnamefont{Vanderbilt}},\ }%
  \bibfield{journal}{%
  \Doi{10.1103/RevModPhys.84.1419}{\bibinfo {journal} {Rev. Mod. Phys.}}\ }%
  \textbf{\bibinfo {volume} {84}},\ \bibinfo {pages} {1419} (\bibinfo {month}
  {Oct}\ \bibinfo {year} {2012})%
  \bibAnnoteFile{NoStop}{MLWF-rmp}%
\bibitem{fu-prb06}%
  \BibitemOpen
  \bibfield{author}{%
  \bibinfo {author} {\bibfnamefont{L.}~\bibnamefont{Fu}}\ and\ \bibinfo
  {author} {\bibfnamefont{C.~L.}\ \bibnamefont{Kane}},\ }%
  \bibfield{journal}{%
  \bibinfo {journal} {Phys. Rev. B}\ }%
  \textbf{\bibinfo {volume} {74}},\ \bibinfo {pages} {195312} (\bibinfo {year}
  {2006})%
  \bibAnnoteFile{NoStop}{fu-prb06}%
\bibitem{soluyanov-prb12}%
  \BibitemOpen
  \bibfield{author}{%
  \bibinfo {author} {\bibfnamefont{A.~A.}\ \bibnamefont{Soluyanov}}\ and\
  \bibinfo {author} {\bibfnamefont{D.}~\bibnamefont{Vanderbilt}},\ }%
  \bibfield{journal}{%
  \Doi{10.1103/PhysRevB.85.115415}{\bibinfo {journal} {Phys. Rev. B}}\ }%
  \textbf{\bibinfo {volume} {85}},\ \bibinfo {pages} {115415} (\bibinfo {month}
  {Mar}\ \bibinfo {year} {2012})%
  \bibAnnoteFile{NoStop}{soluyanov-prb12}%
\bibitem{comment_rashba}%
  \BibitemOpen
  \bibinfo {note} {The $\mathbf{k}$-dependence of the Rashba coupling terms is
  such that they vanish at K and K$'$.}%
  \bibAnnoteFile{Stop}{comment_rashba}%
\bibitem{murakami-prl06}%
  \BibitemOpen
  \bibfield{author}{%
  \bibinfo {author} {\bibfnamefont{S.}~\bibnamefont{Murakami}},\ }%
  \bibfield{journal}{%
  \Doi{10.1103/PhysRevLett.97.236805}{\bibinfo {journal} {Phys. Rev. Lett.}}\
  }%
  \textbf{\bibinfo {volume} {97}},\ \bibinfo {pages} {236805} (\bibinfo {month}
  {Dec}\ \bibinfo {year} {2006})%
  \bibAnnoteFile{NoStop}{murakami-prl06}%
\bibitem{hongbin-prb12}%
  \BibitemOpen
  \bibfield{author}{%
  \bibinfo {author} {\bibfnamefont{H.}~\bibnamefont{Zhang}}, \bibinfo {author}
  {\bibfnamefont{F.}~\bibnamefont{Freimuth}}, \bibinfo {author}
  {\bibfnamefont{G.}~\bibnamefont{Bihlmayer}}, \bibinfo {author}
  {\bibfnamefont{S.}~\bibnamefont{Bl\"ugel}},\ and\ \bibinfo {author}
  {\bibfnamefont{Y.}~\bibnamefont{Mokrousov}},\ }%
  \bibfield{journal}{%
  \Doi{10.1103/PhysRevB.86.035104}{\bibinfo {journal} {Phys. Rev. B}}\ }%
  \textbf{\bibinfo {volume} {86}},\ \bibinfo {pages} {035104} (\bibinfo {month}
  {Jul}\ \bibinfo {year} {2012})%
  \bibAnnoteFile{NoStop}{hongbin-prb12}%
\bibitem{bi-tb}%
  \BibitemOpen
  \bibfield{author}{%
  \bibinfo {author} {\bibfnamefont{Y.}~\bibnamefont{Liu}}\ and\ \bibinfo
  {author} {\bibfnamefont{R.~E.}\ \bibnamefont{Allen}},\ }%
  \bibfield{journal}{%
  \Doi{10.1103/PhysRevB.52.1566}{\bibinfo {journal} {Phys. Rev. B}}\ }%
  \textbf{\bibinfo {volume} {52}},\ \bibinfo {pages} {1566} (\bibinfo {month}
  {Jul}\ \bibinfo {year} {1995})%
  \bibAnnoteFile{NoStop}{bi-tb}%
\bibitem{oh-prl12}%
  \BibitemOpen
  \bibfield{author}{%
  \bibinfo {author} {\bibfnamefont{M.}~\bibnamefont{Brahlek}}, \bibinfo
  {author} {\bibfnamefont{N.}~\bibnamefont{Bansal}}, \bibinfo {author}
  {\bibfnamefont{N.}~\bibnamefont{Koirala}}, \bibinfo {author}
  {\bibfnamefont{S.-Y.}\ \bibnamefont{Xu}}, \bibinfo {author}
  {\bibfnamefont{M.}~\bibnamefont{Neupane}}, \bibinfo {author}
  {\bibfnamefont{C.}~\bibnamefont{Liu}}, \bibinfo {author}
  {\bibfnamefont{M.~Z.}\ \bibnamefont{Hasan}},\ and\ \bibinfo {author}
  {\bibfnamefont{S.}~\bibnamefont{Oh}},\ }%
  \bibfield{journal}{%
  \Doi{10.1103/PhysRevLett.109.186403}{\bibinfo {journal} {Phys. Rev. Lett.}}\
  }%
  \textbf{\bibinfo {volume} {109}},\ \bibinfo {pages} {186403} (\bibinfo
  {month} {Oct}\ \bibinfo {year} {2012})%
  \bibAnnoteFile{NoStop}{oh-prl12}%
\bibitem{armitage1}%
  \BibitemOpen
  \bibfield{author}{%
  \bibinfo {author} {\bibfnamefont{L.}~\bibnamefont{Wu}}, \bibinfo {author}
  {\bibfnamefont{M.}~\bibnamefont{Brahlek}}, \bibinfo {author}
  {\bibfnamefont{R.~V.}\ \bibnamefont{Aguilar}}, \bibinfo {author}
  {\bibfnamefont{A.}~\bibnamefont{Stier}}, \bibinfo {author}
  {\bibfnamefont{C.}~\bibnamefont{Morris}}, \bibinfo {author}
  {\bibfnamefont{Y.}~\bibnamefont{Lubashevsky}}, \bibinfo {author}
  {\bibfnamefont{L.}~\bibnamefont{Bilbro}}, \bibinfo {author}
  {\bibfnamefont{N.}~\bibnamefont{Bansal}}, \bibinfo {author}
  {\bibfnamefont{S.}~\bibnamefont{Oh}},\ and\ \bibinfo {author}
  {\bibfnamefont{N.}~\bibnamefont{Armitage}},\ }%
  \bibfield{journal}{%
  \bibinfo {journal} {Nature Physics}\ }%
  \textbf{\bibinfo {volume} {9}},\ \bibinfo {pages} {410} (\bibinfo {year}
  {2013})%
  \bibAnnoteFile{NoStop}{armitage1}%
\bibitem{jpl-prb13}%
  \BibitemOpen
  \bibfield{author}{%
  \bibinfo {author} {\bibfnamefont{J.}~\bibnamefont{Liu}}\ and\ \bibinfo
  {author} {\bibfnamefont{D.}~\bibnamefont{Vanderbilt}},\ }%
  \bibfield{journal}{%
  \Doi{10.1103/PhysRevB.88.224202}{\bibinfo {journal} {Phys. Rev. B}}\ }%
  \textbf{\bibinfo {volume} {88}},\ \bibinfo {pages} {224202} (\bibinfo {month}
  {Dec}\ \bibinfo {year} {2013})%
  \bibAnnoteFile{NoStop}{jpl-prb13}%
\bibitem{QE-2009}%
  \BibitemOpen
  \bibfield{author}{%
  \bibinfo {author} {\bibfnamefont{P.}~\bibnamefont{Giannozzi}}, \bibinfo
  {author} {\bibfnamefont{S.}~\bibnamefont{Baroni}}, \bibinfo {author}
  {\bibfnamefont{N.}~\bibnamefont{Bonini}}, \bibinfo {author}
  {\bibfnamefont{M.}~\bibnamefont{Calandra}}, \bibinfo {author}
  {\bibfnamefont{R.}~\bibnamefont{Car}}, \bibinfo {author}
  {\bibfnamefont{C.}~\bibnamefont{Cavazzoni}}, \bibinfo {author}
  {\bibfnamefont{D.}~\bibnamefont{Ceresoli}}, \bibinfo {author}
  {\bibfnamefont{G.~L.}\ \bibnamefont{Chiarotti}}, \bibinfo {author}
  {\bibfnamefont{M.}~\bibnamefont{Cococcioni}}, \bibinfo {author}
  {\bibfnamefont{I.}~\bibnamefont{Dabo}}, \bibinfo {author}
  {\bibfnamefont{A.}~\bibnamefont{{Dal Corso}}}, \bibinfo {author}
  {\bibfnamefont{S.}~\bibnamefont{de~Gironcoli}}, \bibinfo {author}
  {\bibfnamefont{S.}~\bibnamefont{Fabris}}, \bibinfo {author}
  {\bibfnamefont{G.}~\bibnamefont{Fratesi}}, \bibinfo {author}
  {\bibfnamefont{R.}~\bibnamefont{Gebauer}}, \bibinfo {author}
  {\bibfnamefont{U.}~\bibnamefont{Gerstmann}}, \bibinfo {author}
  {\bibfnamefont{C.}~\bibnamefont{Gougoussis}}, \bibinfo {author}
  {\bibfnamefont{A.}~\bibnamefont{Kokalj}}, \bibinfo {author}
  {\bibfnamefont{M.}~\bibnamefont{Lazzeri}}, \bibinfo {author}
  {\bibfnamefont{L.}~\bibnamefont{Martin-Samos}}, \bibinfo {author}
  {\bibfnamefont{N.}~\bibnamefont{Marzari}}, \bibinfo {author}
  {\bibfnamefont{F.}~\bibnamefont{Mauri}}, \bibinfo {author}
  {\bibfnamefont{R.}~\bibnamefont{Mazzarello}}, \bibinfo {author}
  {\bibfnamefont{S.}~\bibnamefont{Paolini}}, \bibinfo {author}
  {\bibfnamefont{A.}~\bibnamefont{Pasquarello}}, \bibinfo {author}
  {\bibfnamefont{L.}~\bibnamefont{Paulatto}}, \bibinfo {author}
  {\bibfnamefont{C.}~\bibnamefont{Sbraccia}}, \bibinfo {author}
  {\bibfnamefont{S.}~\bibnamefont{Scandolo}}, \bibinfo {author}
  {\bibfnamefont{G.}~\bibnamefont{Sclauzero}}, \bibinfo {author}
  {\bibfnamefont{A.~P.}\ \bibnamefont{Seitsonen}}, \bibinfo {author}
  {\bibfnamefont{A.}~\bibnamefont{Smogunov}}, \bibinfo {author}
  {\bibfnamefont{P.}~\bibnamefont{Umari}},\ and\ \bibinfo {author}
  {\bibfnamefont{R.~M.}\ \bibnamefont{Wentzcovitch}},\ }%
  \bibfield{journal}{%
  \bibinfo {journal} {Journal of Physics: Condensed Matter}\ }%
  \textbf{\bibinfo {volume} {21}},\ \bibinfo {pages} {395502 (19pp)} (\bibinfo
  {year} {2009})%
  \bibAnnoteFile{NoStop}{QE-2009}%
\bibitem{pbe-1}%
  \BibitemOpen
  \bibfield{author}{%
  \bibinfo {author} {\bibfnamefont{J.~P.}\ \bibnamefont{Perdew}}, \bibinfo
  {author} {\bibfnamefont{K.}~\bibnamefont{Burke}},\ and\ \bibinfo {author}
  {\bibfnamefont{M.}~\bibnamefont{Ernzerhof}},\ }%
  \bibfield{journal}{%
  \Doi{10.1103/PhysRevLett.77.3865}{\bibinfo {journal} {Phys. Rev. Lett.}}\ }%
  \textbf{\bibinfo {volume} {77}},\ \bibinfo {pages} {3865} (\bibinfo {month}
  {Oct}\ \bibinfo {year} {1996})%
  \bibAnnoteFile{NoStop}{pbe-1}%
\bibitem{pbe-2}%
  \BibitemOpen
  \bibfield{author}{%
  \bibinfo {author} {\bibfnamefont{J.~P.}\ \bibnamefont{Perdew}}, \bibinfo
  {author} {\bibfnamefont{K.}~\bibnamefont{Burke}},\ and\ \bibinfo {author}
  {\bibfnamefont{M.}~\bibnamefont{Ernzerhof}},\ }%
  \bibfield{journal}{%
  \Doi{10.1103/PhysRevLett.78.1396}{\bibinfo {journal} {Phys. Rev. Lett.}}\ }%
  \textbf{\bibinfo {volume} {78}},\ \bibinfo {pages} {1396} (\bibinfo {month}
  {Feb}\ \bibinfo {year} {1997})%
  \bibAnnoteFile{NoStop}{pbe-2}%
\bibitem{opium-web}%
  \BibitemOpen
  \bibinfo {howpublished} {\url{http://opium.sourceforge.net/}}%
  \bibAnnoteFile{NoStop}{opium-web}%
\bibitem{opium-paper}%
  \BibitemOpen
  \bibfield{author}{%
  \bibinfo {author} {\bibfnamefont{N.~J.}\ \bibnamefont{Ramer}}\ and\ \bibinfo
  {author} {\bibfnamefont{A.~M.}\ \bibnamefont{Rappe}},\ }%
  \bibfield{journal}{%
  \Doi{10.1103/PhysRevB.59.12471}{\bibinfo {journal} {Phys. Rev. B}}\ }%
  \textbf{\bibinfo {volume} {59}},\ \bibinfo {pages} {12471} (\bibinfo {month}
  {May}\ \bibinfo {year} {1999})%
  \bibAnnoteFile{NoStop}{opium-paper}%
\bibitem{monkhorst-pack}%
  \BibitemOpen
  \bibfield{author}{%
  \bibinfo {author} {\bibfnamefont{H.~J.}\ \bibnamefont{Monkhorst}}\ and\
  \bibinfo {author} {\bibfnamefont{J.~D.}\ \bibnamefont{Pack}},\ }%
  \bibfield{journal}{%
  \Doi{10.1103/PhysRevB.13.5188}{\bibinfo {journal} {Phys. Rev. B}}\ }%
  \textbf{\bibinfo {volume} {13}},\ \bibinfo {pages} {5188} (\bibinfo {month}
  {Jun}\ \bibinfo {year} {1976})%
  \bibAnnoteFile{NoStop}{monkhorst-pack}%
\bibitem{wannier90}%
  \BibitemOpen
  \bibfield{author}{%
  \bibinfo {author} {\bibfnamefont{A.~A.}\ \bibnamefont{Mostofi}}, \bibinfo
  {author} {\bibfnamefont{J.~R.}\ \bibnamefont{Yates}}, \bibinfo {author}
  {\bibfnamefont{Y.~S.}\ \bibnamefont{Lee}}, \bibinfo {author}
  {\bibfnamefont{I.}~\bibnamefont{Souza}}, \bibinfo {author}
  {\bibfnamefont{D.}~\bibnamefont{Vanderbilt}},\ and\ \bibinfo {author}
  {\bibfnamefont{N.}~\bibnamefont{Marzari}},\ }%
  \bibfield{journal}{%
  \Doi{{10.1016/j.cpc.2007.11.016}}{\bibinfo {journal} {Computer Phys. Comm.}}\
  }%
  \textbf{\bibinfo {volume} {178}},\ \bibinfo {pages} {685} (\bibinfo {year}
  {2008})%
  \bibAnnoteFile{NoStop}{wannier90}%
\bibitem{comment_wannier90}%
  \BibitemOpen
  \bibinfo {note} {The TB models from Wannier90 are constructed in such a way
  that the Wannier-interpolated bandstructure reproduces the first-principles
  bandstructure exactly within an energy window centered around the Fermi
  level.}%
  \bibAnnoteFile{Stop}{comment_wannier90}%
\bibitem{fukui-jpsj05}%
  \BibitemOpen
  \bibfield{author}{%
  \bibinfo {author} {\bibfnamefont{T.}~\bibnamefont{Fukui}}, \bibinfo {author}
  {\bibfnamefont{Y.}~\bibnamefont{Hatsugai}},\ and\ \bibinfo {author}
  {\bibfnamefont{H.}~\bibnamefont{Suzuki}},\ }%
  \bibfield{journal}{%
  \bibinfo {journal} {Journal of the Physical Society of Japan}\ }%
  \textbf{\bibinfo {volume} {74}},\ \bibinfo {pages} {1674} (\bibinfo {year}
  {2005})%
  \bibAnnoteFile{NoStop}{fukui-jpsj05}%
\bibitem{soluyanov-prb11}%
  \BibitemOpen
  \bibfield{author}{%
  \bibinfo {author} {\bibfnamefont{A.~A.}\ \bibnamefont{Soluyanov}}\ and\
  \bibinfo {author} {\bibfnamefont{D.}~\bibnamefont{Vanderbilt}},\ }%
  \bibfield{journal}{%
  \Doi{{10.1103/PhysRevB.83.235401}}{\bibinfo {journal} {{Phys. Rev. B}}}\ }%
  \textbf{\bibinfo {volume} {{83}}},\ \bibinfo {pages} {{235401}} (\bibinfo
  {month} {{JUN 2}}\ \bibinfo {year} {{2011}}),\ ISSN \bibinfo {issn}
  {{1098-0121}}%
  \bibAnnoteFile{NoStop}{soluyanov-prb11}%
\end{thebibliography}%

\end{document}